\documentclass[aps,prb,twocolumn,noshowpacs,superscriptaddress,amssymb,amsfonts,amsmath,citeautoscript]{revtex4}
\usepackage{bm}
\usepackage{graphicx}

\usepackage[dvips]{color} 
\begin{document}

\def \ed {\end{document}}
\def \bc {\begin{comment}} \def \ec {end{comment}}
\def\Fbox#1{\vskip1ex\hbox to 8.5cm{\hfil\fboxsep0.3cm\fbox{%
  \parbox{8.0cm}{#1}}\hfil}\vskip1ex\noindent}  


\newcommand{\eq}[1]{(\ref{#1})}
\newcommand{\Eq}[1]{Eq.~(\ref{#1})}
\newcommand{\Eqs}[1]{Eqs.~(\ref{#1})}
\newcommand{\Fig}[1]{Fig.~\ref{#1}}
\newcommand{\Figs}[1]{Figs.~\ref{#1}}
\newcommand{\Sec}[1]{Sec.~\ref{#1}}
\newcommand{\Secs}[1]{Secs.~\ref{#1}}
\newcommand{\Ref}[1]{Ref.~\cite{#1}}
\newcommand{\Refs}[1]{Refs.~\cite{#1}}

\def\be{\begin{equation}}\def\ee{\end{equation}}
\def\bea{\begin{eqnarray}}\def\eea{\end{eqnarray}}
\def\bse{\begin{subequations}}\def\ese{\end{subequations}}

\let \nn  \nonumber  \newcommand{\br}{\\ \nn}
\newcommand{\BR}[1]{\\ \label{#1}}
\def\hf{\frac{1}{2}}
\let \= \equiv \let\*\cdot \let\~\widetilde \let\^\widehat \let\-\overline
\let\p\partial \def\pp {\perp} \def\pl {\parallel}
\def\ort#1{\^{\bf{#1}}}
\def\Trans{^{\scriptscriptstyle{\rm T}}}
\def\x{\ort x} \def\y{\ort y} \def\z{\ort z}
 \def\bn{\bm\nabla} \def\1{\bm1} \def\Tr{{\rm Tr}}
\def\Re{\mbox{  Re}}
\def\<{\left\langle}    \def\>{\right\rangle}
\def\({\left(}          \def\){\right)}
 \def \[ {\left [} \def \] {\right ]}

\renewcommand{\a}{\alpha}\renewcommand{\b}{\beta}\newcommand{\g}{\gamma}
\newcommand{\G} {\Gamma}\renewcommand{\d}{\delta}
\newcommand{\D}{\Delta}\newcommand{\e}{\epsilon}\newcommand{\ve}{\varepsilon}
\newcommand{\E}{\Epsilon}\renewcommand{\o}{\omega} \renewcommand{\O}{\Omega}
\renewcommand{\L}{\Lambda}\renewcommand{\l}{\lambda}
\renewcommand{\t}{\tau}
\def\r{\rho}\def\k{\kappa}
\def\t{\theta } \def\T{\Theta } \def\s{\sigma} \def\S{\Sigma}

\newcommand{\B}[1]{{\bm{#1}}}
\newcommand{\C}[1]{{\mathcal{#1}}}    
\newcommand{\BC}[1]{\bm{\mathcal{#1}}}
\newcommand{\F}[1]{{\mathfrak{#1}}}
\newcommand{\BF}[1]{{\bm{\F {#1}}}}

\renewcommand{\sb}[1]{_{\text {#1}}}  
\renewcommand{\sp}[1]{^{\text {#1}}}  
\newcommand{\Sp}[1]{^{^{\text {#1}}}} 
\def\Sb#1{_{\scriptscriptstyle\rm{#1}}}

\title{Mutual friction in superfluid $^3$He-B in the low-temperature regime}

\author{J.T.~M\"akinen}
\affiliation{Low Temperature Laboratory, Department of Applied Physics, Aalto University, FI-00076 AALTO, Finland}

\author{V.B.~Eltsov}
\affiliation{Low Temperature Laboratory, Department of Applied Physics, Aalto University, FI-00076 AALTO, Finland}


\date{\today}

\begin{abstract}
We measure the response of a rotating sample of superfluid $^3$He-B to spin-down to rest in the zero-temperature limit. Deviations from perfect cylindrical symmetry in the flow environment cause the initial response to become turbulent. The remaining high polarization of vortices along the rotation axis suppresses the turbulent behavior and leads to laminar late-time response. We determine the dissipation during laminar decay at $(0.13-0.22) T_{\mathrm{c}}$ from the precession frequency of the remnant vortex cluster. We extract the mutual friction parameter $\alpha$ and confirm that its dependence on temperature and pressure agrees with theoretical predictions. We find that the zero-temperature extrapolation of $\alpha$ has pressure-independent value $\alpha(T=0) \sim 5 \cdot 10^{-4}$, which we attribute to a process where Kelvin waves, excited at surfaces of the container, propagate into the bulk and enhance energy dissipation via overheating vortex core-bound fermions.
\end{abstract}
%

\maketitle 

\section{Introduction}

The superfluid phases of $^3$He were the first experimentally accessible macroscopic quantum systems where the multi-component order parameter supports a variety of quantized vortices with non-singular cores.\cite{PhysRevLett.51.1362,doublequantum} Such vortices can possess hard cores with radius of the order of the coherence length, filled with a superfluid phase different of that in the bulk. The vortex-core-bound fermions play an important role in the dynamics of the vortices. They interact with the bulk thermal excitations, leading to a force, called mutual friction, acting on a vortex in an external flow field. The mutual friction is well understood at higher temperatures $(T \gtrsim 0.3 T_{\mathrm{c}})$ both theoretically\cite{PhysRevB.44.9667} and experimentally,\cite{bevan_mf,PhysRevLett.105.125301} but the quantitative experimental confirmation in the zero temperature limit has been absent. Furthermore, non-zero extrapolation of dissipation to $T = 0$ has been observed both in $^4$He\cite{PhysRev.136.A1194} and in $^3$He.\cite{Eltsov2015} In $^4$He experiments the remnant dissipation can be attributed to $^3$He impurities in the sample.\cite{PhysRev.136.A1194} On the other hand, superfluid $^3$He is isotopically pure, but finite zero-temperature extrapolation is observed nonetheless.

Spin-down measurements, where a steadily rotating container is abruptly brought to rest, provide well controlled access to superfluid vortex dynamics. During the steady rotation the quantized vortices form a well-defined lattice in which the vortex density is controlled by the angular velocity. When the container is brought to rest, the normal component imposes a force on vortices. In $^4$He the post-spin-down dynamics are always turbulent,\cite{Walmsley2008,PhysRevLett.99.265302} while in $^3$He in a cylindrical container the response is found to be laminar at least down to $0.20$~$T_\mathrm{c}$.\cite{PhysRevLett.105.125301} Deviations from perfect cylindrical symmetry\cite{PhysRevB.85.224526} or introduction of dissipative AB phase boundary\cite{PhysRevB.84.184532} lead to (at least partially) turbulent response to spin-down also in $^3$He-B.

In turbulent spin-down the dissipation is greatly enhanced by vortex reconnections,\cite{PhysRevB.85.224526,PhysRevB.84.184532,PhysRevLett.99.265302} in particular. The resulting time scales are generally much faster than for laminar vortex motion, where the scale is determined by mutual friction. In this work we apply nuclear magnetic resonance (NMR) techniques \cite{Eltsov2011} and Andreev scattering of quasiparticles\cite{Blazkova2007,Bradley2009,Blaauwgeers2007,PhysRevB.85.224526} to probe the vortex dynamics after spin-down to rest in $^3$He-B in the $T \rightarrow 0$ limit.

%

\section{Mutual friction}

Vortex motion $\mathbf{r} = \mathbf{r}_{0} + (\mathbf{v}_{\mathrm{L}} - \mathbf{v}_{\mathrm{n}})t$ with respect to the normal fluid motion $\mathbf{v}_{\mathrm{n}}$ leads to pumping of core-bound fermions along the zero-crossing branch in the energy spectrum.\cite{0295-5075-32-8-006} This phenomenon is known as spectral flow. Here $\mathbf{v}_{\mathrm{L}}$ is the velocity of the vortex, $\mathbf{r}_{0} = \mathbf{r}(t=0)$ its initial position, and $t$ is time. The energy levels of the core-bound fermions are separated approximately by energy $\hbar \omega_{0} \sim \Delta^{2}/E_{\mathrm{F}} \ll \Delta$, called the minigap. Here $E_{\mathrm{F}}$ is the Fermi-energy and $\Delta$ is the superfluid gap. Relaxation of the core-bound fermions towards the thermal equilibrium with time constant $\tau$ leads to a net force acting on a vortex\cite{Donnelly,RevModPhys.59.87,doi:10.1080/00018736000101169}
\begin{equation}
 \mathbf{F}_{\mathrm{N}} = D(\mathbf{v}_{\mathrm{n}}-\mathbf{v}_{\mathrm{L}})_{\bot} + D'\mathbf{\hat{z}} \times (\mathbf{v}_{\mathrm{n}}-\mathbf{v}_{\mathrm{L}}).
\end{equation}
Parameters $D$ and $D'$ are given by\cite{PhysRevB.44.9667}
\begin{equation}
 D = \rho \kappa \frac{\omega_{0} \tau}{1+\omega_{0}^{2} \tau^{2}} \tanh \frac{\Delta(T)}{2 k_{\mathrm{B}}T}
\end{equation}
and
\begin{equation}
 D' = \rho \kappa \left[ 1 - \frac{\omega_{0}^{2} \tau^{2}}{1 + \omega_{0}^{2} \tau^{2}} \tanh
 \frac{\Delta(T)}{2 k_{\mathrm{B}}T} \right] - \rho_{s} \kappa,
\end{equation}
where $\tau$ is the average lifetime of Bogoliubov quasiparticles at the Fermi surface,\cite{einzel.1978} $-\rho_{\mathrm{s}}\kappa$ is the Iordanskii force,\cite{IORDANSKY1964335,PhysRevLett.79.1321}  $\rho=\rho_{\mathrm{s}}+\rho_{\mathrm{n}}$ is the total fluid density where $\rho_{\mathrm{s}}$ and $\rho_{\mathrm{n}}$ are the densities of the superfluid and the normal components, respectively, and $T$ is temperature.

If the vortex mass is neglected, the total force acting on a vortex, which includes mutual friction and Magnus forces, should be zero. The force balance can be written as\cite{bevan_mf}
\begin{equation}
 (\mathbf{v}_{\mathrm{n}}-\mathbf{v}_{\mathrm{L}})\times \hat{\mathbf{z}} + \alpha(\mathbf{v}_{\mathrm{n}}-\mathbf{v}_{\mathrm{s}})_{\bot} + \alpha'\mathbf{\hat{z}} \times (\mathbf{v}_{\mathrm{n}}-\mathbf{v}_{\mathrm{s}}) = 0,
\end{equation}
where the first term is the Magnus force. The mutual friction parameters $\alpha$ and $\alpha'$ are defined as
\begin{equation}
 \alpha = \frac{D/\kappa \rho_{\mathrm{s}}}{(D/\kappa \rho_{\mathrm{s}})^{2} + (1 - D'/\kappa \rho_{\mathrm{s}})^{2}}
\end{equation}
and
\begin{equation}
 \alpha' = 1 - \frac{1 - D'/\kappa \rho_{\mathrm{s}}}{(D/\kappa \rho_{\mathrm{s}})^{2} + (1 - D'/\kappa \rho_{\mathrm{s}})^{2}}.
\end{equation}
In the $T\rightarrow 0$ limit the reactive parameter $\alpha' \sim \alpha^{2}$ can safely be neglected. In the same limit the dissipative term becomes
\begin{equation}\label{eq:alphaLT}
 \alpha \sim \frac{1}{\omega_{0} \tau}
\end{equation}
and the temperature dependence is dominated by the quasiparticle lifetime
\begin{equation}
 \tau \propto \exp \left( \frac{\Delta}{k_{\mathrm{B}} T} \right).
\end{equation}


\section{Laminar superflow and response to spin-down}

The coarse-grained hydrodynamic Hall-Vinen-Bekarevich-Khalatnikov equation for the superfluid velocity $\mathbf{v}_{\mathrm{s}}$ is\cite{Donnelly}
\begin{equation} \label{eq:coarse-grained}
 \frac{\partial \mathbf{v}_{\mathrm{s}}}{\partial t} + \nabla \mu - (\mathbf{v}_{\mathrm{s}} \cdot \nabla)\mathbf{v}_{\mathrm{s}} =
 -\alpha \hat{\omega} \times [( \mathbf{v}_{\mathrm{s}} - \mathbf{v}_{\mathrm{n}}) \times (\nabla \times \mathbf{v}_{\mathrm{s}})],
\end{equation}
where $\mu$ is the chemical potential and $\hat{\omega}$ is a unit vector along the vorticity. Assuming that vortices remain highly polarized along the rotational axis as generally is the case for rotating superflow,\cite{PhysRevLett.115.155303,ncommhosio} vortex reconnections can be ignored. The superfluid mimics laminar solid-body-like motion and quantized vortices create combined superfluid velocity field $\mathbf{v}_{\mathrm{s}} = \Omega_{\mathrm{s}}\mathbf{\hat{z}} \times \mathbf{r}$, where $\Omega_{\mathrm{s}}$ is the angular velocity of rotation around axis $\mathbf{\hat{z}}$. With this form of $\mathbf{v}_{\mathrm{s}}$ Eq.~\eqref{eq:coarse-grained} transforms after taking curl of both sides to
\begin{equation} \label{eq:laminar}
 \frac{\mathrm{d}\Omega_{\mathrm{s}}(t)}{\mathrm{d}t} = 2 \alpha \Omega_{\mathrm{s}}(t) [\Omega - \Omega_{\mathrm{s}}(t)],
\end{equation}
where $\Omega$ is the angular velocity of the normal component, assumed to be equal to the drive. If a step-like change from $\Omega = \Omega_{0}$ to 0 is performed at $t=0$, the response at $t>0$ follows
\begin{equation} \label{eq:omega_s}
 \Omega_{\mathrm{s}}(t) = \frac{\Omega_{0}}{1 + t/\tau},
\end{equation}
where $\tau = (2 \alpha \Omega_{0})^{-1}$. In reality, the step is performed at finite rate $-\dot{\Omega}$. During the deceleration, i.e., for $-\Omega_{0}/\dot{\Omega} < t < 0$, Eq.~(\ref{eq:laminar}) has solution
\begin{equation}
 \Omega_{\mathrm{s}}(t) = \frac{\sqrt{\dot{\Omega}}\exp[\alpha(t+\Omega_{0}/\dot{\Omega})(\Omega_{0}-\dot{\Omega}t)]}
 {\tau_{0}\sqrt{\dot{\Omega}} + \sqrt{\pi \alpha} \exp(\alpha \Omega_{0}^{2}/\dot{\Omega})\mathrm{erf}
 \left( \sqrt{\alpha \dot{\Omega}}t \right) },
\end{equation}
where $\tau_{0} = \Omega_{0}^{-1} + \sqrt{\alpha/\dot{\Omega}}\exp(\alpha \Omega_{0}^{2}/\dot{\Omega}) \mathrm{erf} \left( \sqrt{\alpha/\dot{\Omega}}\Omega_{0} \right)$.

At low temperatures the typical time scales of the vortex dynamics are much longer than those of the deceleration in our experiments, i.e. $\alpha \ll \dot{\Omega}/\Omega_{\mathrm{0}}^{2}$, and at the end of the deceleration $\Omega_{\mathrm{s}}(t=0) \cong \Omega_{\mathrm{0}}$. It is thus justified to use Eq.~(\ref{eq:omega_s}) at all times $t > 0$.

\section{Experimental setup}

\begin{figure}
\includegraphics[width=0.3\textwidth]{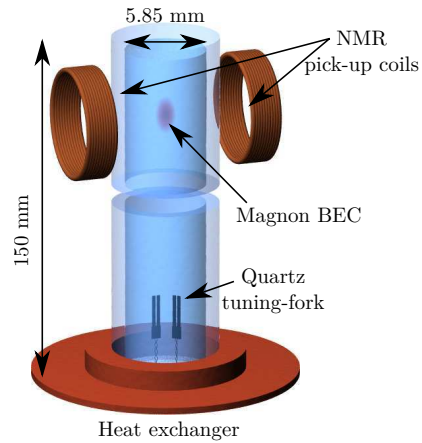}
\caption{\label{fig:setup}
Experimental setup used in the measurements. The container (not to scale) is a quartz-glass cylinder with smooth walls to avoid vortex pinning. It is rotated about its vertical axis. The NMR pick-up coils, located close to the top of the container, are used to probe the vortex dynamics along with two quartz tuning forks located at the bottom close to the heat exchanger. The quartz tuning forks are additionally used for thermometry since they are sensitive probes for local quasiparticle density. The bottom of the container is open to a heat exchanger volume with rough surfaces covered with sintered silver.}
\end{figure}

The $^3$He-B sample is contained in a 150 mm long smooth-walled cylindrical quartz-glass container with $5.85$~mm inner diameter, illustrated in Fig.~\ref{fig:setup}. The bottom of the container is open to silver-sintered surface acting as a heat exchanger. The pressure in the sample is varied between 0 and 29 bar and the sample can be cooled down to $0.13 T_{\mathrm{c}}$. The sample is rotated with angular velocities up to $2$~rad/s. The maximum rate of deceleration is $-\dot{\Omega} = -0.03$~rad/s$^2$. The axial symmetry is broken by two quartz tuning forks, used as thermometers,\cite{Blazkova2007,Bradley2009,Blaauwgeers2007} and by vortex pinning to the sintered surface at the bottom. The inner surfaces of the quartz glass cylinder were treated with hydrofluoric acid\cite{cell_treatment} to avoid vortex pinning elsewhere.

In the ballistic regime the forks' resonance width is proportional to the Boltzmann factor $\exp (-\Delta/k_{B}T)$.\cite{Blazkova2007} In the presence of a superfluid flow field, created for example by a nearby vortex bundle, the forks' resonance width becomes a function of the surrounding vortex structure.\cite{PhysRevLett.93.235302,PhysRevB.85.224526} Owing to Andreev reflection the vortices shadow part of the heat flow emanating from the walls of the container.\cite{PhysRevB.96.054510} After a spin-down we see oscillations with increasing period in the resonance width of the fork, see Fig.~\ref{fig:fork_example}. We interpret these oscillations as caused by a precessing vortex cluster which develops some rotational asymmetry as a result of the spin-down.

Our setup also includes a set of NMR pick-up coils, used to probe the spatial distribution of the order-parameter, called texture.\cite{Heikkinen2013} We apply rf pulses that excite transverse spin waves, or magnon quasiparticles. Pumped magnons quickly form Bose-Einstein condensate (BEC) in the magneto-textural trap close to the axis of the sample.\cite{PhysRevLett.108.145303,PhysRevLett.69.3092} The competing effect of the axial magnetic field and of the boundary conditions for the order parameter imposes smooth variation to the order parameter in the radial direction, forming an effective potential well for magnons. In the axial direction magnons are trapped by a shallow minimum in the magnetic field, created by an external solenoid. The textural part of the trap is modified in the presence of vortices due to contributions from their cores and the associated superfluid velocity field.\cite{Eltsov2011} Magnetization of the magnon BEC coherently precesses with a frequency which depends on the trapping potential. Therefore, the NMR measurements allow us to probe the evolution of vortex distribution within the trap by periodic application of excitation pulses. After a spin-down to rest the measured frequency of coherent precession oscillates with increasing period, see Fig.~\ref{fig:magnon_example}. This observation further supports the interpretation about a precessing nonuniform vortex cluster after the spin-down.

\begin{figure}
\includegraphics[width=.45\textwidth]{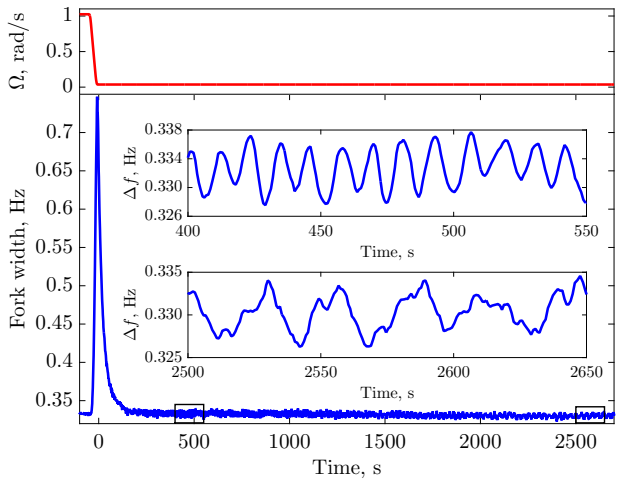}
\caption{\label{fig:fork_example}
Response of the resonance width of the thermometer fork recorded during spin-down from $\Omega_{0} = 1.02$~rad/s to rest. The initial overshoot is caused by heat produced by turbulent dissipation of the vortex cluster after the spin-down. Time $t=0$ corresponds to the moment when the drive $\Omega$ reaches zero. The insets show zoomed view of the late-time response. The periodic oscillations originate from precession of a remnant vortex cluster. The increase of the oscillation period with time is used to extract the dissipation.}
\end{figure}

In the measurements the period $p(t)$, extracted from temporal separation of the local maxima in the NMR or fork response, is converted into the angular velocity of the precessing vortex cluster using
\begin{equation} \label{eq:period}
 \Omega_{\mathrm{s}}(t) = \frac{2 \pi}{p(t)}.
\end{equation}
Here we assume that the local maximum of either type of signal is related to position of some identifiable feature in the precessing vortex cluster and thus the temporal separation of two subsequent maxima corresponds to a single round of vortex precession in the container.

\section{Measurements on laminar decay of precessing vortex cluster}

The initial vortex density is controlled by the angular velocity $\Omega_{\mathrm{0}}$, so that the aerial density of vortices is equal to solid-body-rotation value $n_{\mathrm{v}}=2\Omega_{\mathrm{0}} /\kappa$. To prepare the initial state the sample is rotated at velocity $\Omega > \Omega_{0}$ before returning to $\Omega_{0}$ to ensure enough vortices are created. In some measurements this step is done at about $0.7 T_{\mathrm{c}}$, where vortex dynamics is fast. Afterwards the sample is cooled down to the desired temperature over a time period of the order of an hour. Alternatively, similar procedure is done at lower temperatures. In this case steady rotation at $\Omega_{0}$ is maintained for a few hours before the spin-down. We ensure that the dissipation of the magnon BEC, proportional to vortex density, has reached constant value before the spin-down. After the spin-down the response $\Omega_{\mathrm{s}}(t)$ is monitored for as long as the oscillations are seen, typically for a few hours.

A short turbulent burst, seen as an initial overshoot of the fork width\cite{Eltsov2015} in Fig.~\ref{fig:fork_example}, is observed as soon as the deceleration starts. The first oscillations are typically seen right after the turbulent $t^{-3/2}$ decay of vortex line density, some $\sim 100$~s after the container is at rest. We use the initial angular velocity $\Omega_{\mathrm{s}}(t=0) \equiv \Omega_{\mathrm{i}}$ and the time constant $\tau$ as fitting parameters in Eq.~(\ref{eq:omega_s}). We find that $\Omega_{\mathrm{i}} \sim ( 0.6-0.8 ) \Omega_{\mathrm{0}}$ in all our measurements. Thus, we estimate that $20-40\%$ of vortices are lost during the initial turbulent burst. The mutual friction parameter $\alpha$ is extracted from
\begin{equation} \label{eq:alpha}
 \alpha = \frac{1}{2 \Omega_{\mathrm{i}} \tau}
\end{equation}
as a function of temperature at three different pressures, see Fig. \ref{fig:alpha_vs_df}. We find that $\alpha$ has, within the accuracy of our measurements, a linear dependence on the width of the quartz tuning fork as expected in the $T \rightarrow 0$ limit. The measurements also show a finite pressure-independent zero-temperature extrapolation $\alpha_{0} \equiv \alpha(T \rightarrow 0)$. In the ballistic regime $\alpha$ can be written as a function of the fork resonance width as
\begin{equation} \label{eq:forkwidth}
\alpha = \alpha_{0} + B \Delta f = \alpha_{0} + B C \exp(-\Delta/k_{\mathrm{B}}T),
\end{equation}
where the coefficient $B$ is the slope in Fig.~\ref{fig:alpha_vs_df}. Parameter $C$ is a geometrical factor specific to the type of the resonator, which in our case has been determined to have the value $C = 10.0 \pm 1.5$~kHz at 0.5 bar pressure by calibrating the fork against the Leggett frequency at $0.37$~$T_{\mathrm{c}}$ in $^3$He-B.\cite{Thuneberg2001} The geometrical factor scales as $C \propto p_{F}^{4}$ as a function of pressure.\cite{Bradley2009} We compare the measured value $B \Delta f$ with the expected behavior $\sim (\omega_{0} \tau)^{-1}$ as a function of pressure. The results are shown in Fig.~\ref{fig:prefactor_vs_pressure}. We use low temperature minigap values from Ref.~\citenum{PhysRevLett.115.235301}, interpolated in $\Delta_{0}^{2} p_{\mathrm{F}}^{-1}$ using quadratic fit. The results show the expected pressure dependence. While the absolute value agrees with earlier measurements at 29 bar pressure,\cite{PhysRevB.84.184532} its magnitude is a factor of $6$ smaller than the value of $(\omega_{0} \tau)^{-1}$.

\begin{figure}
\includegraphics[width=.465\textwidth]{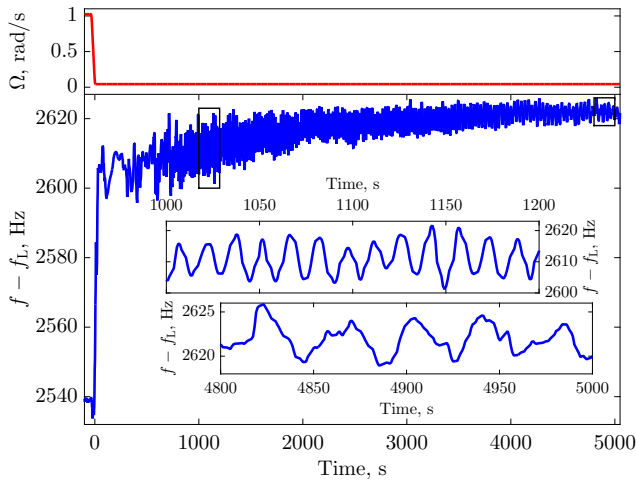}
\caption{\label{fig:magnon_example}
Temporal evolution of the ground state of magnon BEC after a spin-down from $\Omega_{0} = 1.02$~rad/s to rest. The initial increase in the frequency is caused by the drop in the vortex density and decreased polarization during the  turbulent burst. The insets show zoomed view of late-time behavior, where oscillations in the ground state are caused by periodic modulation of spatial distribution of the order parameter by a precessing vortex cluster.}
\end{figure}

We use the weak-coupling-plus bulk gap with strong coupling correction. We have found no need for the gap renormalization, contrary to the fits at $T>0.3 T_{\mathrm{c}}$ presented in Ref.~\citenum{bevan_mf}. Thus, we believe that the measured values of $\alpha$ can be directly compared with the theory.

\section{Possible sources of finite friction at $T \rightarrow 0$}

Finite dissipation in quantum turbulence in the zero temperature limit has previously been observed in superfluid $^4$He,\cite{PhysRevLett.99.265302,Davis200043} and in $^3$He-B.\cite{PhysRevLett.96.035301,PhysRevLett.99.265301,Eltsov2015} The microscopic sources of dissipation, as well as the role of the normal component, are quite different for the two superfluids. In superfluid $^4$He the normal component has independent dynamics, which couples to the dynamics of the superfluid component via mutual friction. At large drives the dynamics of normal component in $^4$He may be turbulent. Non-zero density of the normal component and thus friction may exist even in the $T \rightarrow 0$ limit when $^3$He impurities are present. Otherwise, the zero-temperature dissipation is believed to originate from acoustic emission by rapidly oscillating vortices,\cite{Vinenturb} which terminates the Kelvin-wave cascade. So far the experimental verification of this scenario is absent. In $^3$He-B the normal component is practically always laminar and its density vanishes exponentially towards lower temperatures. Here we consider a few possible dissipation mechanisms as candidates for the observed zero-temperature dissipation in laminar motion in $^3$He-B.

\begin{figure}
\includegraphics[width=.45\textwidth]{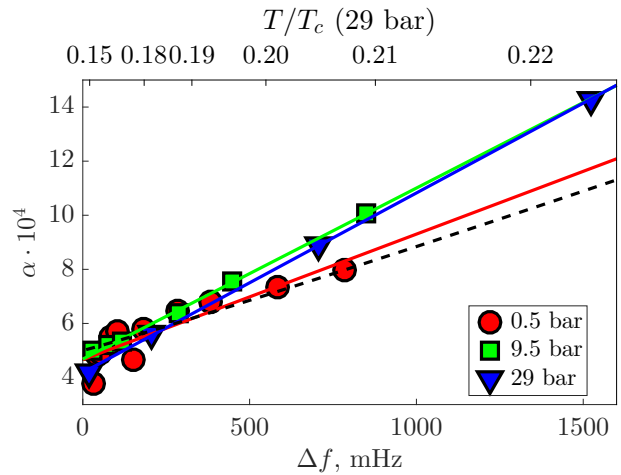}
\caption{\label{fig:alpha_vs_df}
Dissipative mutual friction parameter $\alpha$ as a function of the quartz tuning fork resonance width $\Delta f$. The dashed line follows Eq.~\eqref{eq:forkwidth} at 0.5 bar pressure, assuming $\alpha_{0} = 5 \cdot 10^{-4}$ and $B\Delta f = (6\omega_{0} \tau)^{-1}$, and the solid lines are fits to the same equation at different pressures. The fork width is converted to $T/T_{\mathrm{c}}$ scale at 29 bar at the top axis.}
\end{figure}

One possibility is surface friction in the presence of rough surfaces like the silver-sintered ones in the heat exchanger. The authors in Ref.~\citenum{PhysRevB.84.184532} studied the response of $^3$He-B to spin-down by measuring the magnitude of counterflow in a cylinder with smooth walls and possibility to introduce a slab with high dissipation. The region with high dissipation could be created in the middle of the sample by using magnetic field to stabilize a layer of superfluid $^3$He-A. At low temperatures the mutual friction coefficients in the A-phase are orders of magnitude larger than those in the B-phase. In the presence of the A-B phase boundary the response in the B-phase was always turbulent. Additionally, the flow profile during the decay was clearly different from solid-body like.

In the absence of the phase boundary laminar behavior was observed down to the lowest measured temperature $0.20$~$T_{\mathrm{c}}$, with $\Omega_{\mathrm{i}} \tau = 740$. According to Eq. (\ref{eq:alpha}) this corresponds to $\alpha \simeq 7 \cdot 10^{-4}$, which is in good agreement with our current work. This observation suggests that the surface friction can not be accounted for by simple increase of the mutual friction coefficient $\alpha$ but it leads to qualitatively different behavior.

Consider a vortex moving along a dissipative surface. The energy dissipation from the motion is limited by the vortex tension $T_{\mathrm{v}} = \kappa^{2} \rho_{\mathrm{s}}(4 \pi)^{-1} \ln \frac{b}{a}$, where $b$ is the intervortex separation and $a$ is the vortex core size. The existence of this limit has been previously observed previously in spin-up measurements on $^4$He.\cite{PhysRevB.32.171}

\begin{figure}
\includegraphics[width=.45\textwidth]{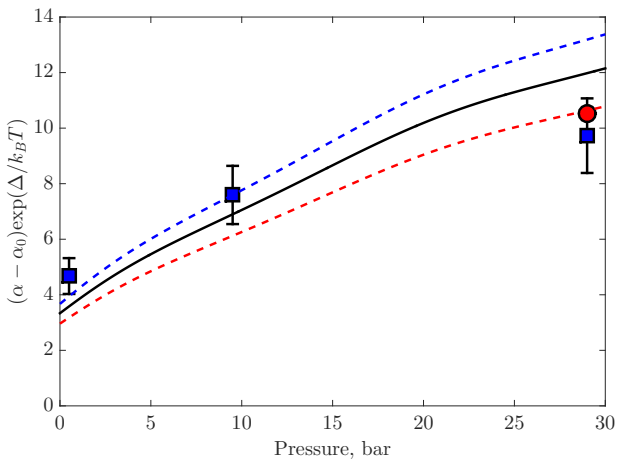}
\caption{\label{fig:prefactor_vs_pressure}Coefficient $(\alpha-\alpha_{0})\exp(\Delta/k_{B}T)$ as a function of pressure. Blue squares correspond to measurements in this work and the red circle is extracted from data in Ref.~\citenum{PhysRevB.84.184532}, measured at $0.20$~$T_{\mathrm{c}}$, assuming the same geometrical factor $C$ and $\alpha_{0} = 5 \cdot 10^{-4}$. Error bars correspond to the inaccuracy of the determination of the geometrical factor $C$ in Eq.~\eqref{eq:forkwidth}. The lines follow $(6 \omega_{0}\tau)^{-1}$ and correspond to $(0.20,0.165,$ and $0.13)T_{\mathrm{c}}$ from top to bottom, respectively.}
\end{figure}

Assuming maximum pulling force, the surface dissipation power can be calculated as
\begin{equation} \label{eq:dissip_power}
 W = \int_{0}^{R} T_{\mathrm{v}} n_{\mathrm{v}} |v_{\mathrm{ex}}(r)| 2 \pi r \mathrm{d}r = \frac{1}{3} \kappa \rho_{\mathrm{s}} \Omega_{\mathrm{s}}^{2} R^{3} \ln \frac{b}{a}.
\end{equation}
Here $n_{\mathrm{v}} = 2\Omega_{\mathrm{s}}/\kappa$ is the vortex density and $v_{\mathrm{ex}} = \Omega_{\mathrm{s}}r$ is velocity of the vortex ends relative to the surface.

Using Eqs.~\eqref{eq:laminar} and \eqref{eq:dissip_power} we can write an equation for decay of total kinetic energy $E$ after a step-like spin-down
\begin{equation} \label{eq:diff_for_e}
 \frac{\mathrm{d}E}{\mathrm{d}t} = -\beta \frac{4 \kappa \ln \frac{b}{a}}{3 \pi R H} E - \frac{8 \alpha_{\mathrm{sfm}}}{\sqrt{M R^{2}}} E^{3/2}.
\end{equation}
Here $E = M (R \Omega_{\mathrm{s}})^{2}$, $M = \pi R^{2} H \rho_{\mathrm{s}}$ is the mass of the superfluid in the container, $H$ is the height of the container, and $0 \leq \beta \leq 1$ is a dimensionless parameter describing the effectiveness of the surface friction. The first term in Eq.~\eqref{eq:diff_for_e} describes the surface dissipation and the second term mutual friction in the bulk. We extract $E(t)$ from the experiments using $\Omega_{\mathrm{s}}(t)$ from Eq.~\eqref{eq:period} and use the initial condition $E(t=0) = M (R \Omega_{\mathrm{sfm}})^{2}$. Parameters $\alpha_{\mathrm{sfm}}$, $\beta$, and $\Omega_{\mathrm{sfm}}$ are used as fitting parameters. The subscript {\it sfm} refers to {\it surface friction model}, i.e. to the fitted parameter value in this model. We find $\beta \ll 1$, while $\alpha_{\mathrm{sfm}}$ and $\Omega_{\mathrm{sfm}}$ agree with the previously fitted $\alpha$ and $\Omega_{i}$. Thus, the model including the surface friction, Eq.~\eqref{eq:dissip_power}, when applied to our data effectively reduces to the one without, Eq.~\eqref{eq:omega_s}. Alternatively, we tried fixing $\alpha_{\mathrm{sfm}}$ to $\alpha(\Delta f)-\alpha_{0}$, where $\alpha(\Delta f)$ is taken from Eq.~\eqref{eq:laminar}. We find that we can not reproduce the observed decay of $\Omega_{\mathrm{s}}$ this way, see Fig.~\ref{fig:example_decay}. The finite zero-temperature dissipation $\alpha_{0}$ can not thus be replaced by surface friction.

\begin{figure}
\includegraphics[width=.45\textwidth]{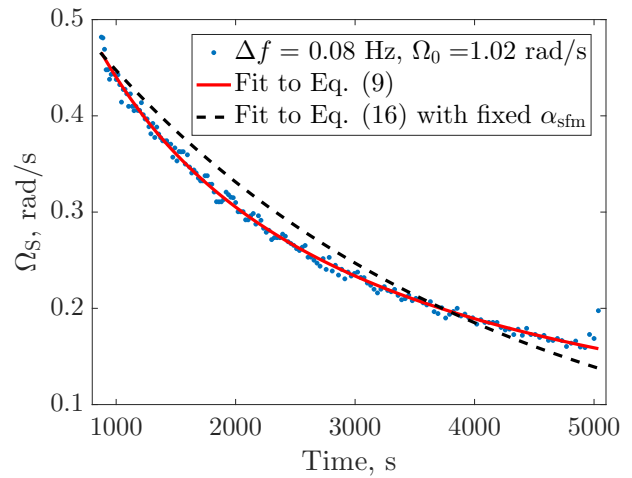}
\caption{\label{fig:example_decay}The dots show an example decay at 9.5~bar pressure using data from Fig.~\ref{fig:magnon_example}. The solid red line is a fit to laminar decay model, Eq.~\eqref{eq:laminar}, which only includes mutual friction. The dashed black line shows the behavior of the extended model, Eq.~\eqref{eq:diff_for_e}, with fixed $\alpha_{\mathrm{sfm}}$ and fitted $\beta$, see details in the text.}
\end{figure}

Another possible source for finite zero-temperature dissipation is proposed by Silaev in Ref.~\citenum{PhysRevLett.108.045303}. In this model, the vortex-core-bound fermions interact with the flow, obtain additional energy, and escape the core if the vortex is in accelerating motion. The process is effective down to the absolute zero temperature. There are at least two sources of accelerating vortex motion in our measurements, whose effects we will now consider.  

The first source for acceleration is the centrifugal motion of the remnant vortex cluster around the rotation axis after the spin-down. Consider a cluster of vortices moving with angular velocity $\Omega_{\mathrm{s}}$ with respect to the container. The vortex-core-bound fermions approximately follow heat balance equation\cite{PhysRevLett.108.045303}
\begin{equation} \label{eq:silaev_fric}
 \mathcal{A} \xi /v_{\mathrm{c}}^{3} = \frac{\Delta_{0}}{k_{\mathrm{B}} T_{\mathrm{loc}}}\exp \left(-\frac{\Delta_{0}}{k_{\mathrm{B}}T_{\mathrm{loc}}}\right),
\end{equation}
where $\mathcal{A}= \langle u_{\mathrm{x}} \dot{u}_{\mathrm{y}} - u_{\mathrm{y}} \dot{u}_{\mathrm{x}} \rangle_{t}/2$, $\xi$ is the coherence length, $\Delta_{0}$ is the superfluid energy gap at zero temperature, $v_{\mathrm{c}} = \Delta_{0}p_{\mathrm{F}}^{-1}$ is the superfluid critical velocity, and $T_{\mathrm{loc}}$ is the temperature inside the vortex core. The brackets denote time average and $u_{\mathrm{x}}$ and $u_{\mathrm{y}}$ are the $x$ and $y$ components of the vortex velocity $u(t)$, respectively. For circular periodic motion at distance $R$ from the axis, we get
\begin{equation}
 \mathcal{A} \sim R^{2} \Omega_{\mathrm{s}}^{3}.
\end{equation}
We estimate the dissipation by substituting $29$~bar parameter values at $T=0$ and using typical experimental values $\Omega_{\mathrm{s}} = 1$~rad/s and $R=3$~mm. We get $k_{\mathrm{B}} T_{\mathrm{loc}}/\Delta_{0} \approx 0.02$, which is equal to $T_{\mathrm{loc}} \sim 0.04$~$T_{\mathrm{c}}$. The dissipation caused by the centrifugal motion thus only dominates the vortex dynamics below $0.04 T_{\mathrm{c}}$, which is much lower than the lowest  temperatures in our measurements. It seems unlikely that the centrifugal motion is responsible for the extra dissipation.

The second source for acceleration is the presence of vortex waves, such as Kelvin waves. Estimation for $\mathcal{A}$ for typical Kelvin waves was done by Silaev in Ref.~\citenum{PhysRevLett.108.045303}. According to this estimate the presence Kelvin waves overheats the cores to temperature $T_{\mathrm{loc}} \sim 0.2 T_{\mathrm{c}}$. We note that $\alpha_{0} \approx \alpha(T=0.2 T_{\mathrm{c}})-\alpha(T=0)$. In other words, the unaccounted dissipation at zero temperature corresponds to temperature increase of roughly $0.2 T_{\mathrm{c}}$, in agreement with Silaev's estimate for $T_{\mathrm{loc}}$. The Kelvin waves are naturally expected to be created immediately after the spin-down during the initial turbulent burst.\cite{PhysRevB.92.184508} However, the dissipation related to Kelvin waves should decrease and eventually cease as the initial small scale vortex waves decay. Laminar vortex motion is not a likely candidate for providing energy input to small scales since there are no vortex reconnections.

There is, however, at least one persistent source of Kelvin waves. We suggest that vortex motion along a rough surface, such as that of the heat exchanger, generates Kelvin waves that then propagate along the vortices. In principle, one expects that generation of Kelvin waves is more prominent in areas where vortex motion with respect to the surface is faster, i.e. at larger distance from the container axis. We note, however, that vortex waves have been seen to transfer to nearby vortices.\cite{PhysRevFluids.1.084501} Eventually this process can bring the whole volume to a quasi-uniform state with all vortices having similar Kelvin-wave spectrum. Generation of Kelvin waves from vortex motion along a dissipative surface could thus effectively lead to enhanced dissipation in the whole volume via Silaev's mechanism.


\section{Conclusions}

We have studied the spin-down response of superfluid $^3$He-B in a cylindrical container at $T=(0.13-0.22)$~$T_{\mathrm{c}}$ at $0.5, 9.5,$ and $29$~bar pressures. Deviations from cylindrical symmetry in our setup lead to an initial turbulent burst, followed by hours-long laminar decay. We extract mutual friction parameter $\alpha$ from the evolution of angular velocity of the remnant vortex cluster. We find that $\alpha$ depends exponentially on the temperature as theoretically expected. The observed pressure dependence is explained by the behavior of the minigap separating the energy levels of the vortex-core-bound fermions. The absolute values of the friction coefficient are a factor of 6 lower than the theoretical estimation.

The zero-temperature extrapolation of $\alpha$ reveals a pressure-independent finite value $\alpha_{0} \approx 5 \cdot 10^{-4}$. We consider surface friction and a mechanism proposed by Silaev in Ref.~\citenum{PhysRevLett.108.045303} as possible sources for zero-temperature dissipation. The latter requires that vortices are in accelerating motion. We rule out surface friction and Silaev friction from precessing motion of vortices as possible sources of the observed extra dissipation, while oscillating vortex motion from sufficiently developed Kelvin waves could provide enough dissipation. The Kelvin waves produced in an initial turbulent burst after the spin-down decay during laminar motion at later times. Thus they could not support time-independent $\alpha_{0}$ seen in our observations. We propose that vortex motion along the rough surface of the heat exchanger at the bottom of the experimental container generates Kelvin waves, which propagate into the bulk along the vortices effectively enhancing dissipation in the whole volume via mechanism proposed by Silaev.

In future it would be interesting to measure mutual friction in a system where vortex interactions with boundaries can be neglected. One possibility may, in principle, be provided by freely propagating vortex rings, which can be created in the experiments i.e. by a moving grid.\cite{gridturb} One should be careful, though, that even in such system Kelvin waves, excited on the rings at the moment of formation, can significantly affect further dynamics.\cite{vortexrings}


\begin{acknowledgments}
We thank M. A. Silaev for useful discussions. This work was supported by the Academy of Finland (grant no. 284594, LTQ CoE). Our research made use of the OtaNano – Low Temperature Laboratory infrastructure of Aalto University, that is part of the European Microkelvin Platform.
\end{acknowledgments}


\end{document}